\documentclass[conference]{IEEEtran}
\IEEEoverridecommandlockouts
\usepackage{cite}
\usepackage[fleqn]{amsmath}
\usepackage{amssymb,amsfonts}
\usepackage{graphicx}
\usepackage{textcomp}
\usepackage{xcolor}
\usepackage{url}
\usepackage[noend]{algpseudocode}
\usepackage{algorithmicx,algorithm}
\usepackage{color}
\usepackage{subfigure}
\usepackage{multirow}
\usepackage{makecell}
\def\BibTeX{{\rm B\kern-.05em{\sc i\kern-.025em b}\kern-.08em
    T\kern-.1667em\lower.7ex\hbox{E}\kern-.125emX}}
\begin{document}

\title{AI-oriented Medical Workload Allocation for Hierarchical Cloud/Edge/Device Computing}

%\author{Tianshu Hao \and
%Jianfeng Zhan \thanks{Jianfeng Zhan is the corresponding author.}\and
%Wanling Gao \and
%Xu Wen \and
%Kai Hwang
%}
%
%\institute{State Key Laboratory of Computer Architecture, Institute of Computing Technology, Chinese Academy of Sciences\\
%\email{\{haotianshu, zhanjianfeng, gaowanling, wenxu\}@ict.ac.cn}
%}

\author{
\IEEEauthorblockN{Tianshu Hao\IEEEauthorrefmark{1}\IEEEauthorrefmark{2},  Jianfeng Zhan\IEEEauthorrefmark{1}\IEEEauthorrefmark{2}\thanks{Jianfeng Zhan is the corresponding author.}, Kai Hwang\IEEEauthorrefmark{3}\IEEEauthorrefmark{4}, Wanling Gao\IEEEauthorrefmark{1}\IEEEauthorrefmark{2}, Xu Wen\IEEEauthorrefmark{1}\IEEEauthorrefmark{2}}
  
\IEEEauthorblockA{\IEEEauthorrefmark{1}State Key Laboratory of Computer Architecture, Institute of Computing Technology, Chinese Academy of Sciences\\
\{haotianshu, zhanjianfeng, gaowanling, wenxu\}@ict.ac.cn
}

\IEEEauthorblockA{\IEEEauthorrefmark{3} The Chinese University of Hong Kong, Shenzhen, China, hwangkai@cuhk.edu.cn}
\IEEEauthorblockA{\IEEEauthorrefmark{2} University of Chinese Academy of Sciences} 
\IEEEauthorblockA{\IEEEauthorrefmark{4} Shenzhen Institute of Artificial Intelligence and Robotics for Society} 
}

\maketitle

\begin{abstract}
In a hierarchically-structured cloud/edge/device computing environment, workload allocation can greatly affect the overall system performance. This paper deals with AI-oriented medical workload generated in emergency rooms (ER) or intensive care units (ICU) in metropolitan areas. The goal is to optimize AI-workload allocation to cloud clusters, edge servers, and end devices so that minimum response time can be achieved in life-saving emergency applications.

In particular, we developed a new workload allocation method for the AI workload in distributed cloud/edge/device computing systems. An efficient scheduling and allocation strategy is developed in order to reduce the overall response time to satisfy multi-patient demands. We apply several ICU AI workloads from a comprehensive edge computing benchmark Edge AIBench. The healthcare AI applications involved are short-of-breath alerts, patient phenotype classification, and life-death threats. Our experimental results demonstrate the high efficiency and effectiveness in real-life health-care and emergency applications.
\end{abstract}


\section{Introduction}
With the rapid development of user-end Internet of Things (IoT)~\cite{Devices,hwang2017big}, edge computing emphasizes real-time computing near the end devices~\cite{shi2016promise}. A common hierarchical framework of distributed cloud and edge computing consists of three layers: cloud cluster, edge server, and end devices. The principal advantage of this distributed computing paradigm is reducing the latency by reducing the time of data transmission to users~\cite{satyanarayanan2017emergence}.

Artificial Intelligence (AI) technology is widely used in edge computing to support edge intelligence~\cite{wang2019edge}. There are a lot of edge AI scenarios closed to human daily life, such as smart home, smart health, smart factory and so on. What's more, most of them are latency-sensitive applications. Thus how to reduce the response time of these edge AI applications has become a really important problem.

Many corporations, such as Google's edge TPU~\cite{cass2019taking} and Intel's neural compute stick~\cite{di2018embedded}, put efforts on edge AI accelerator chips to speed up the processing time on the edge device. And~\cite{jia2014caffe,tflite} implements the light-weighted model to reduce the complexity of the AI models which can be applied to edge servers. And there is also some benchmarking work to evaluate the inference of AI workload on edge devices. \cite{luo2018aiot,ignatov2018ai} have evaluated the inference performance on the edge devices. 

However, common practices usually deploy real-time processing on edge servers. But they don't consider changing the workload deployment layer for different workloads. Different allocation strategies of the AI workload greatly influence the response time. Deploying the inference process of an AI workload on the edge server layer doesn't always achieve the minimum latency. Therefore, an AI workload allocation model is lacked for edge computing latency-sensitive applications now. Considering different device computational abilities of each level and peak network bandwidth, how to make the trade-off between them is still a problem in edge computing hierarchical framework. 

We extract medical AI workloads as the problem environment from Edge AIBench~\cite{hao2018edge} in this paper. Based on the hierarchically-structured framework, this paper proposes a workload allocation strategy for the medical edge AI workload. And we give a numerical interpretation of the response time of a single workload. What's more, we also propose an efficient scheduling algorithm for multi-job AI scenarios to reduce the total response time of jobs.

We attempt to evaluate the application response time in the hierarchical computing system in the following two aspects.

1) Processing time: Since the AI workloads are usually compute-intensive applications, the processing time is decided by the computing ability function.

2) Transmission time: The data transmission time is represented by the network function.

The workload allocation aims to reduce the response time for the medical latency-sensitive applications. We give a latency reduction algorithm for single and multiple workloads. Finally, we use the real-world edge AI workload and ICU datasets from Edge AIBench~\cite{hao2018edge} to validate the efficiency and effectiveness. The experimental results show our allocation algorithm can get the minimum response time comparing with other strategies.

The rest of this paper is organized as follows. Section II introduces the problem environment of the hierarchically-structured framework and edge AI workload. Section III and IV present the workload allocation problem and an optimal algorithm for a single job. In section V and VI, we present an efficient allocation and scheduling algorithm for multi jobs. Section VII gives the experimental setup and section VIII shows the experiment results analysis. Section IX introduces recent and related work. Finally, we conclude in section X.

\section{The Problem Environment}
\subsection{Hierarchically-structured Cloud/Edge/Device Computing}

\begin{figure}[ht]
\centering
\includegraphics[scale=0.4]{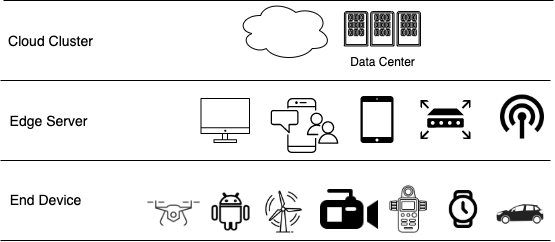}
\caption{Hierarchical Framework of Edge Computing.}
\label{framework}
\end{figure}

Edge computing is a distributed computing paradigm with the three-layer framework: cloud cluster, edge computing server and user-side end devices. Figure~\ref{framework} shows the hierarchically-structured cloud/edge/device computing framework of edge computing.

The cloud cluster is a centralized server cluster remote away from the edge devices. Traditionally in cloud computing, most of the computing tasks are being executed in the datacenter. But the transmission latency from the cloud server to user-side devices is long, sometimes even longer than the processing time. Therefore, in the edge computing framework, the cloud server usually executes offline tasks and central control.

Edge computing layer server is closer to users than the cloud cluster, which can be personal computers, mobile phones, routers, base stations and so on. And the edge server has more computational resources than user-side end devices. Online tasks are usually executed on the edge computing layer.

There is a great diversity of user-side end devices, such as smartwatches, sensors, unmanned aerial vehicles, and smart vehicles, and etc. The end devices usually gather data and conduct some simple preprocess tasks because of the restriction of the computing resources.

Generally in the three-layer framework, the higher the layer, the more computational resources of the device, the faster the speed of processing, but the longer the data transmission time. Thus, how to trade off the computational ability and data location becomes a problem.

\subsection{Medical AI Workloads Benchmark}

Edge AIBench~\cite{hao2018edge} is a comprehensive end-to-end benchmark towards AI edge computing. By surveying the edge computing scenarios extensively, Edge AIBench finds that most of these workloads use AI technology. Therefore, we focus on edge AI workloads in this paper. 

\begin{figure}[ht]
\centering
\includegraphics[scale=0.5]{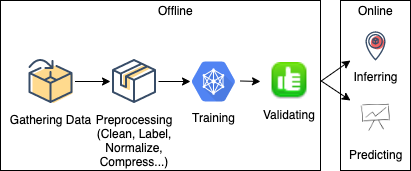}
\caption{AI Online and Offline Workflow.}
\label{AI}
\end{figure}

Figure~\ref{AI} shows the main workflow of AI workloads, including offline and online processes~\cite{muller2016future}. 

Considering the complexity of the AI workload, there are high requirements for computing machines~\cite{chen2018cognitive}. Because the cloud cluster has a better computational ability, offline processes in figure~\ref{AI} are always executed on the cloud cluster in the edge computing framework. And the pre-trained model will be sent to the online processing device after completing training on the cloud cluster. And the most common edge AI workload allocation method is putting the training on the cloud server and putting the inference or other online processes on the edge computing layer~\cite{rahmani2018exploiting, hosseini2017deep}. 

However, where to deploy the online process greatly influence the response time considering the data transmission time. What's more, many edge AI workloads are latency-sensitive applications. Therefore, response time reduction is important in edge computing workload distribution.

From 4 typical edge AI scenarios and 8 application benchmarks in Edge AIBench, we extract the ICU patient monitor as the workload allocation problem background.

\begin{figure}[ht]
\centering
\includegraphics[scale=0.35]{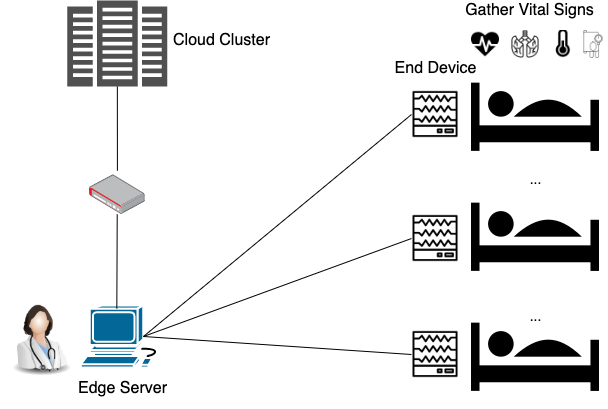}
\caption{ICU Patient Monitor Edge Computing Scenario in Hospital.}
\label{ICU}
\end{figure}

ICU Patient monitor is a typical latency-sensitive edge AI scenario. Because ICU is the treatment place for critical patients, the response latency of any task is very important for doctors' further action. As figure~\ref{ICU} shows, there are several patients in serious condition in an ICU room. Each patient has many end devices to measure vital signs such as temperature, respiration, pulse, and heartbeat, and etc. And there is one end computing device for one patient and one edge server of one ICU room. What's more, there is a cloud server for heavy computation tasks and data storage.

In this scenario, an AI model LSTM~\cite{olah2015understanding} is usually used as the AI network model and real-world ICU patients' data are usually text data.

There are many applications in the ICU scenario and we choose three typical AI workloads from Edge AIBench: short-of-breath alerts, patient phenotype classification, and life-death prediction to conduct the evaluation experiments. Based on the purpose of these workloads, we set different priorities for them.

\section{Latency Analysis for Single Workload}
\subsection{Basic Assumptions}
As figure \ref{framework} shows, the hierarchically-structured framework consists of three levels: $CC$ indicates cloud clusters, $ES$ indicates edge computing servers, and $ED$ indicates end devices. The three major factors that determine the latency of the inference process are the computational ability of each level, the size and the complexity of the model and the dataset, and the network latency. In this paper we make the following assumptions:

(a). The data is gathered from the end devices, thus data transmission needn't be considered if the process is on the end devices.

(b). The transmission time from the cloud cluster to the end device $T_{CC-ED}$ is equal to the sum of the transmission time from the cloud cluster to the edge server $T_{CC-ES}$ plus the transmission time from the edge server to the device $T_{ES-ED}$.

(c). Considering the AI workload is compute-intensive, the computational ability of each layer is expressed by the floating point operations per second (FLOPS)\cite{flops,williams2009roofline} of devices.

(d). To simplify the problem, there are just one cloud server and one edge server to be considered in this model.

(e). The response time of any workload depends on the workload model complexity, dataset size, the device processing the workload, and the network condition.

(f). The transmission time of the inference result $T_{rt}$ is relatively much lower than the transmission time of the inference data $T_{dt}$. Therefore we don't consider $T_{rt}$ in this problem.

Based on the above assumptions, the problem can be expressed as: there are a cloud server, an edge server, and an end device constructed following the hierarchically structured framework as figure \ref{framework} shows. The problem is which layer to allocate the given online AI workload (such as inference) can get the minimum response time.

\begin{figure}[ht]
\flushleft 
\subfigure[Inference on the cloud cluster.]{
\begin{minipage}[t]{1\linewidth}
\centering
\includegraphics[width=2.5in]{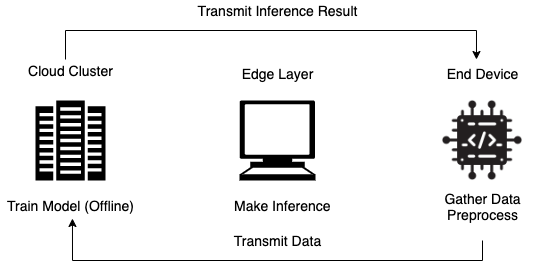}
\end{minipage}
}

\subfigure[Inference on the edge server.]{
\begin{minipage}[t]{1\linewidth}
\centering
\includegraphics[width=2.5in]{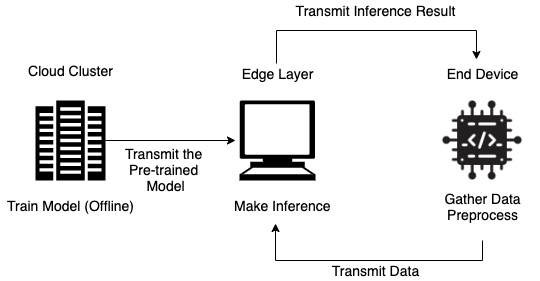}
\end{minipage}
}

\subfigure[Inference on the end device.]{
\begin{minipage}[t]{1\linewidth}
\centering
\includegraphics[width=2.5in]{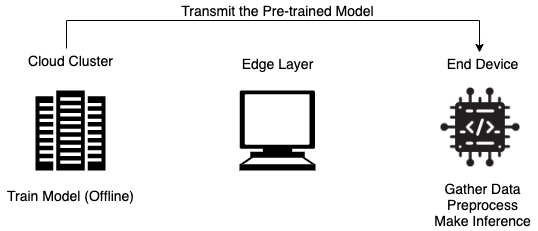}
\end{minipage}
}
\caption{Trade-off among three processing layers of a hierarchically-structured cloud/edge/device environment.}

\label{location}
\end{figure}

From the online tasks in figure~\ref{AI}, we mainly consider the inference allocation problem in this paper for the actual production environment. Figure~\ref{location} shows the three different inference location methods respectively on cloud cluster, edge server, and end devices.

\subsection{Objective Function for Latency Reduction}
In order to find the optimal solution for this problem, we express it as a numerical problem. Table \ref{notation1} shows the notations in this problem.
\begin{table}[htbp]
	\centering
	\caption{Notations of Workload Allocation Problem.} 
	\label{notation1} 
	\begin{tabular}{|p{1.5cm}|p{6cm}|}  
	   \hline
	    Notation&Interpretation \\
		\hline
		$D_i$&the transmission time of data from end device to the execution layer i\\ 
		\hline
		$I_i$&the time of making the inference on the layer i\\
		\hline
		$s$&the size of the workload data\\
		\hline
		$comp$&the FLOPs of the AI model in workload i \\
		\hline
		$AI_i$& the FLOPS of the layer i\\
		\hline
		$\lambda1$&the weight coefficient of the data transmission time\\
		\hline
		$\lambda2$&the weight coefficient of the inference processing time \\
		\hline
	\end{tabular}
\end{table}

For different layers, the probability of whether to make inference on this layer is expressed by:
\begin{align}
p_{i}\in {\{0,1\}}(i \in{\{CC, ES, ED}\})  \nonumber
\\where \quad \sum p_i = 1
\end{align}
The data transimission time $D$ is expressed by:
\begin{align}
D =\lambda_{1}s\sum P_{i} D_{i}
\end{align}

The computational ability of the device is expressed by $AI_{i} = FLOPS  (i \in{\{CC, ES, ED}\}) $.

The inference processing time $I$ is expressed by:
\begin{align}
I =\lambda_{2}\sum P_{i} I_{i},\quad where \quad I_{i} =\frac{s\times comp}{AI_{i}} 
\end{align}

Where the inference processing time on layer i is expressed by $I_{i}$.

Therefore, the objective function can be expressed by:
\begin{align}
minimize \quad T= D+I 
\\
subject \quad to \quad (1)-(3)\nonumber
\end{align}

\subsection{Performance Metrics for AI Workload}
We measure the computational ability of devices from each layer by FLOPS~\cite{flops,williams2009roofline}.  And the FLOPS is calculated by the number of cores $\times$ operating frequency  $\times$ operations per cycle.

And we also measure the complexity of the AI model by FLOPs. The FLOPs is computed using the number of parameters multiplying with the size of the feature map for convolution kernels. The computation formula is $FLOPs = 2HW(C_{in}K^2 + 1)C_{out} $. And for fully connected layers, the FLOPS equals to the number of the parameters~\cite{molchanov2016pruning}. The computation formula is $FLOPs = (2I-1)O$. $H$ indicates the height, $W$ indicates the width and $C_{in}$ indicates the number of channels of the input feature map. $K$ indicates the kernel width and $C_{out}$ indicates the number of output channels. $I$ indicates the input dimensionality and $O$ indicates the output dimensionality\cite{molchanov2016pruning}. 

\section{Allocation Algorithm for Latency Reduction}
This section gives an optimization strategy to get the minimum latency with a given workload and devices environment. In order to solve the problem we defined, we simplify the problem by assuming that the quantity of each layer's device is one. 

For a given edge AI workload and device environment, we can calculate the computational ability of three layers and the network condition firstly. Then we can calculate the minimum response time by putting the inference in different layers. Algorithm 1 shows the procedure of our optimization strategy. 

Firstly, for each workload, we analyze their computation model parameters and calculate the FLOPs $comp$ needed to execute. 

Secondly, we calculate the unit network transmission latency $D_{i_u}$ for deploying on the cloud server and edge server using a unit size extracting from the given dataset. Therefore we can calculate the given workload transmission time $D_i$ by multiplying the dataset size and weight coefficient. 

Thirdly, we calculate the computation ability $AI_i$ of the device in each layer by FLOPS. Because most of the AI workloads are computation-intensity, so we just consider FLOPS here.

The next step is to calculate the weight coefficient for processing time and transmission time. We conduct an experiment to compute the time of one respectively small dataset and get $\lambda_{1}$ and $\lambda_{2}$ by comparing them.

Next, we get the estimated response time for deploying on each layer by summing up the processing time $I_i$ and transmission time $D_i$. 

Finally, we choose the minimum response time layer as the deployment layer for this workload. 

\begin{algorithm}[t]
\caption{Algorithm of Latency Optimization} 
\hspace*{0.02in} {\bf Input:} 
$W$, $s$, $i$, $D_i$\\
\hspace*{0.02in} {\bf Output:}
the minimum response time $T_{min}$
\begin{algorithmic}[1]
\State Calculate the number of FLOPs of the AI workload model $comp$
\State Calculate the unit network latency 
\For{i = CC, ES} 
　　\State $D_{i_u}$ = latency time of unit dataset transmission
\EndFor
\State Calculate the computational of each layer's device
\For{i = CC, ES, ED} 
　　\State  $AI_i = FLOPS\quad of \quad device\quad i$
\EndFor
\State Normalize the transmission time and inference time by calculating the weight coefficient $\lambda_{1}$, $\lambda_{2}$
\State Calculate the inference processing time.
\For{i = CC, ES, ED} 
      \State $I_{i} = \frac{\lambda_{2}\times s \times comp}{AI_{i}}$ 
\EndFor
\State Calculate the data transmission time.
\For{i = CC, ES} 
       \State $D_{i} = \lambda_{1}*s*D_{i_u}$
\EndFor
\State Calculate the minimum latency
\State Let minimum response time $T_{min}= +\infty $
\For( $i = CC, ES, ED$)
      \State $p_i=1$
      \State  $T_i = I_{i} + D_{i}$
     \If{$T_i < T_min$}
　　　　\State $T_{min} = T_i$
　　\EndIf
\EndFor
\State \Return $T_{min}$
\end{algorithmic}
\end{algorithm}

\section{Workload Allocation for Multiple Jobs}
\subsection{Multi-job Workload}
Considering the real-life ICU edge computing environment, every patient has an end device to conduct preprocessing in one emergency room. And these patients share one edge server in the emergency room and one cloud server remote from the ICU. These machines, including one cloud server, one edge server, and several patients' end devices, can be considered as several unrelated parallel machines.

Each patient's end device may release an inference job randomly. Assume these jobs are released in a time sequence, our goal is to minimize the overall response time of all jobs. Then this problem can be considered as an unrelated parallel machine scheduling problem\cite{zheng2016two,al2019optimize}.

From the above section, we can calculate the response time for every single job deployed on different layers and get the best layer of the minimum latency. And we set the following constraints in this problem:

C1. Each device can only execute one job at a time to simplify the problem.

C2. The preemption of jobs is not allowed in our system.

C3. The release time and response time are normalized as the non-zero integer units of time.

C4. The job data can be transmitted to the execution layer first and wait for executing.

C5. The workload is prioritized by its importance in real-life. The workload with the higher priority needs to be considered firstly.

Thus, the scheduling problem can be summarized as follows: there are n patients' jobs ($T_1, T_2, ... , T_n$) needed to be executed in one emergency room. These jobs have their own priority ($w_1,w_2, ... , w_n$). The bigger the $w_i$, the higher the priority of the job $i$. And these jobs can be processed on one cloud server ($M_c$), one edge server ($M_e$), or an individual patient end device ($M_d$). Each job $T_i$ has a known release time ($R_1, R_2, ... , R_n$) in the time sequence. And the response latency time of job $T_i$ executed on machine $M_j$ is $L_{ij}$. Table \ref{notation2} shows the notation of this problem.

The main goal of this problem is to reduce the whole response time of multiple jobs.

\begin{table}[htbp]
	\centering
	\caption{Notations of Allocation Problem for Multiple Jobs.} 
	\label{notation2} 
	\begin{tabular}{|p{1.5cm}|p{6cm}|}  
	   \hline
	    Notation&Interpretation \\
		\hline
		i&the index of the job, i = 1, 2, ... , n \\ 
		\hline
		j&the index of the machine, j = c (cloud), e (edge), d (device) \\
		\hline
		T&the unit of time, T = 1, 2, ... , $+\infty$\\
		\hline
		$w_i$&the priority weight of job i \\
		\hline
		$P_{ij}$&$P_{ij}$ equals 1 if the job i is processed on the machine j, otherwise equals 0\\
		\hline
		$I_{ijT}$&$I_{ijT}$ equals 1 if the job i is processed on the machine j at time T, otherwise equals 0\\
		\hline
		$L_{i}$&the response time of the job i\\
		\hline
		$L^*_{i}$&the response time of the job i considering the weight \\
		\hline
		$R_i$&the release time of job i\\
		\hline
		$S_i$&the start processing time of job i\\
		\hline
		$E_i$&the completion time of job i\\
		\hline
		$L_{Sum}$&the whole response time of all jobs\\
		\hline
		$E_{last}$&the completion time of the last job\\
		\hline
	\end{tabular}
\end{table}

\subsection{Objective Function to Minimize Reponse Time}
To get the minimum whole response time of all the jobs, the objective function can be expressed by:
\begin{align}
minimize \quad L_{sum} && \nonumber\
\\
where \quad L_{sum} = \sum_{1}^{n}w_i(L_i-R_i) \label{e3}
\end{align}

The whole response time is calculated by summing up the response time of all jobs. The response time $L_i$ of job $i$ equals the end time $E_i$ minus the release time $R_i$.  And for different priorities of these jobs, we multiply the response time with the priority weight $W_i$ to get the new response time $L_{i}^*$. End devices don't need to subject to this constraint because we assume every job has its own end device.

And $p_{ij}$ and $I_{ijT}$ are binary variables equal 0 or 1. The constraint $\sum_{j=1}^{m}p_{ij}=1$ needs to be considered to ensure the job i is exactly only on one machine. And the constraint $\sum_{i=1}^{n}I_{ijT} \leq 1$ ensures there is only one job executing on the machine $j$ at a time.

From section III we know the execution time of job $i$ consists of the transmission time $D_i$ and processing time $I_i$. The exact processing time on the machine is $I_i$. And the job can transmit data to the machine $j$ while another job is running on the same machine. 

\section{Workload Allocation Heuristic Algorithm}
We can change the lower bound from 0 to the sum of the minimum execution time of all the jobs. The lower bound can be expressed by:
\begin{align}
L_{lb} =  \sum_{i}^{n} \min w_i(I_i+D_i)  
\end{align}

Therefore, the problem can be simplified to be selecting the minimum response time of the last release job.

Because the scheduling problem for n jobs executed on m machines is very complicated, so we develop a heuristic greedy algorithm to solve the problem in this section. Considering each patient has its own end device, so the end device is not the shared machine. And jobs can be executed on end devices at the same time.

We can obtain the initial feasible solutions by ensuring the earliest released job to have the shortest response time. And then we optimize the solution by a neighborhood search method~\cite{mladenovic1997variable}. Algorithm 2 shows the concrete steps. And algorithm 2 needs to follow the constraints we mentioned in the previous section.

Firstly, we use algorithm 1 to calculate the estimated execution time of each job deployed on each layer. And then we normalize the response time to the integer time units. We use a heuristic greedy method to get the initial deployment strategy. We find the optimal deployment machine for each job to have the minimum completion time by time sequence. Then we obtain the initial feasible solution.

Then we generate the neighborhood solutions from the current solution by swapping the current job $i$ to another machine. And then calculate the deployment machine of other jobs using the above heuristic greedy method. If the whole response time reduces, then we swap the deployment machine of job $i$.

And we set max number of iterations $maxCount$ as the stopping condition for the algorithm.

\begin{algorithm}[t]
\caption{Multi-job Allocation Heuristic Algorithm} 
\hspace*{0.02in} {\bf Input:} 
$W_i$, $R_i$\\
\hspace*{0.02in} {\bf Output:}
the minimum response time of the whole jobs $L_{sum}$
\begin{algorithmic}[1]
\State Calculate the execution time $L_ij$ of job $i$ deployed on machine $j$. And get the response time matrix $\mathbf{l}$.
\State Get $\mathbf{l^*}$ by multiplying with the job priority $w_i$.
\State $L^*_{ij} = w_i L_{ij}$
\State Normalize matrix $\mathbf{l}^*$.
\State Set $L_{sum} = \sum_{i}^{n} \min L_i$
\State Set $L_{sum}^* = \sum_{i}^{n} w_i\min L_i$
\State Initial the tabu array, 0 represents the job or machine can be switched and 1 represents can't be switched.
\State Set $tabu_m[cc, es, ed] = 0$
\State Optimize the solution by swaping the job to another machine reduce the whole response time.
\State Set maxCount = $C_{max}$(large enough)
\While{maxCount}
   \State Set $tabu_j[1,...,n] = 0$
   \For{i = 1, 2, ... , n}
   \State Set $tabu_m[cc, es, ed] = 0$
   \State from all the jobs which $tab_j[k]=0$ choose the earliest completion job k 
   \State Set $tab_j[k]=1$
   \State Initial the max response time improvement.
   \State Set $V_{max} = 0$
    \For{j = cc, es, ed}
    \If{$tabu_m[j]=0$}
    \State Calculate the $L_{sum}^*$ reduction $V_{ij}$ when swap job i to machine j
      \State $tabu_m[j]=1$
　　\If{$V_{ij} \geq V_{max}$}
       \State $V_{max} = V_{ij}$
       \State $swap=j$
       \EndIf
       \EndIf
        \EndFor
      \If{$V_{max} > 0$}
      \State Swap the job $k$ to machine $swap$
       \EndIf
      \EndFor
     \State $MaxCount=Maxcount-1$
\EndWhile
\State \Return $L_{sum}$
\end{algorithmic}
\end{algorithm}

\section{Experimental Setup}
\subsection{Experimental Environment}
We set the experimental environment to simulate the realistic cloud/edge/device computing scenario. We consider that the cloud server has the highest computational ability, while the edge server's performance is lower and the end device's performance is lowest. Meanwhile, the cloud server has a longer distance to the end device than the edge server. 

We conduct our experiment on three devices respectively representing cloud server, edge server, and end device. The cloud server has 12 2.20-GHz Intel(R) Xeon(R) Gold 5220 CPU cores and 128GB of DDR4 RAM. The edge server has 4 2.20-GHz Intel(R) Xeon(R) Gold 5220 CPU cores and 32GB of DDR4 RAM.  The end device is a Raspberry Pi 4B with 4 1.5-GHz speed Quad-core Broadcom CPU cores and 4GB of DDR4 RAM.

And we refer to~\cite{zhou2017efficient} to set the network latency between the cloud server to the end device as 42ms and network bandwidth as 2.9MB/s. What's more, we measure the network latency between the edge server and end device in our lab LAN environment as 0.239ms and bandwidth as 10MB/s.

Then we calculate the FLOPS of each device using the processor information. Table~\ref{device} shows the basic computational ability of devices of each layer. 

\begin{table}[htbp]
	\centering
	\caption{ Computational Ability of Device on Each Layer.} 
	\label{device} 
	\begin{tabular}{|p{1.5cm}|c|c|c|}  
	   \hline
	    Layer&CPU Cores&CPU Frequency&FLOPS\\
	    \hline
	    Cloud Server&12&2.2GHz&422.4GFLOPS\\
	    \hline
	    Edge Server&4&2.2GHz&140.8GFLOPS\\
		\hline
		End Device&4&1.5GHz&96GFLOPS\\
		\hline
	\end{tabular}
\end{table}

\subsection{AI Workload in ICU Scenario}
For the medical dataset, we choose MIMIC-III~\cite{johnson2016mimic} as the real-world dataset. MIMIC-III includes vital signs and other medical information for over 60000 ICU stays of 40000 unique ICU patients. And we preprocess the original data from the MIMIC-III website~\cite{mimic3web} to get the training data.

We choose three ICU AI applications from Edge AIBench\cite{hao2018edge} for the experiment: short-of-breath alerts, patient phenotype classification, and life-death prediction.

\textbf{Short-of-breath alerts} uses the information of ICU vital signs including glucose, heart rate, height, mean blood pressure, and oxygen, and etc. The purpose of this application is to predict if the patient will suffer short-of-breath later using the LSTM model. Therefore the priority of this application needs to be set high. For the workload of short-of-breath alerts, we set the weight $w$ as 2. And using the number of parameters of the LSTM model in this application, we get the number of FLOPs of this model is 105089.

\textbf{Life-death prediction} uses the information of ICU vital signs including heart rate, height, mean blood pressure, oxygen, and other physiological records. The purpose of this application is to predict whether the patient will die in the hospital. The priority of this application also needs to be set high. So we also set the priority weight $w$ of this application as 2. And using the number of parameters of the LSTM model in this application, we get the number of FLOPs of this model is 7569.

\textbf{Patient phenotype classification} uses the information of the complete ICU physiological record until the time. The purpose of this application is to conduct a 25 separate binary classification task using the LSTM model. Because it's not a very emergency task comparing with the above two tasks. We set lower priority weight $w$ of this application as 1. And using the number of parameters of the LSTM model in this application, we get the number of FLOPs of this model is 347417.

We implement these three ICU AI applications by Python using Tensorflow and Keras~\cite{harutyunyan2017multitask, Harutyunyan2019, gulli2017deep}. And we train these three models offline on our cloud server to get the pre-trained model for inference.

What's more, we set 6 different inference data sizes for ICU AI applications to get 18 different workloads. Table~\ref{workload} shows the concrete information of each workload.

The size of data in table~\ref{workload} is calculated in proportion of the number of record files, the real sizes of these workloads datasets are [700, 1300, 2300, 5000, 10700, 21500, 479, 950, 1900, 3900, 7800, 15900, 836, 1700, 2900, 5300, 10800, 21600] KB.

\begin{table}[htbp]
	\centering
	\caption{AI  Workload Characteristics.} 
	\label{workload} 
	\begin{tabular}{|p{1cm}|p{3cm}|c|c|}  
	   \hline
	    Workload No.&ICU Application&Data Size&Model FLOPs\\
	    \hline
	    WL1-1&Short-of-breath alerts&64&105089\\
	    \hline
	  	    WL1-2&Short-of-breath alerts&128&105089\\
	    \hline
	    	    WL1-3&Short-of-breath alerts&256&105089\\
	    \hline
	    	   WL1-4&Short-of-breath alerts&512&105089\\
	    \hline
	    	    WL1-5&Short-of-breath alerts&1024&105089\\
	    \hline
	    	    WL1-6&Short-of-breath alerts&2048&105089\\
	    \hline
               WL2-1&Life-death prediction&64&7569\\
	    \hline
	          WL2-2&Life-death prediction&128&7569\\
	    \hline
	    WL2-3&Life-death prediction&256&7569\\
	    \hline
	    WL2-4&Life-death prediction&512&7569\\
	    \hline
	    WL2-5&Life-death prediction&1024&7569\\
	    \hline
	    WL2-6&Life-death prediction&2048&7569\\
	    \hline
	    WL3-1&Patient phenotype classification&64&347417\\
	    \hline
	    WL3-2&Patient phenotype classification&128&347417\\
	    \hline
	    WL3-3&Patient phenotype classification&256&347417\\
	    \hline
	    WL3-4&Patient phenotype classification&512&347417\\
	    \hline
	    WL3-5&Patient phenotype classification&1024&347417\\
	    \hline
	    WL3-6&Patient phenotype classification&2048&347417\\
	    \hline
	\end{tabular}
\end{table}

\section{Experiments and Performance Analysis}
\subsection{Single Medical Workload Allocation}
Firstly for each workload, we calculate the estimated response time of deploying on different layers by using algorithm 1. Then we choose the optimal deployment layer referring to the results. Table~\ref{estimated} shows our estimated computation results and the best deployment layer for each workload. 

\begin{table}[htbp]
	\centering
	\caption{Estimated Response Time Using Algorithm 1 at the Cloud, Edge, and Device Levels.} 
	\label{estimated} 
	\begin{tabular}{|c|p{1.6cm}|c|c|c|}  
	   \hline
	   \multicolumn{1}{|p{0.8cm}|} {\multirow{2}{*}{ \shortstack{Worklo-\\ad No.}}}
	       &   
	    \multicolumn{1}{p{1.6cm}|}{\multirow{2}{*}{ \shortstack{Chosen Deplo- \\yment Layer}}}
	   &
	   \multicolumn{3}{c|} {Estimated Response Time for Deploying on} \\
	    \cline{3-5}&&Cloud Server&Edge Server&End Device
	\\
	    \hline
	    WL1-1&Edge Server&2091&1279&1394\\
	    \hline
	  	    WL1-2&Edge Server&4182&2558&2788\\
	    \hline
	    	    WL1-3&Edge Server&8364&5116&5576\\
	    \hline
	    	   WL1-4&Edge Server&16728&10232&11152\\
	    \hline
	    	    WL1-5&Edge Server&33456&20464&22304\\
	    \hline
	    	    WL1-6&Edge Server&66912&40928&44608\\
	    \hline
               WL2-1&End Device&212&109&79\\
	    \hline
	          WL2-2&End Device&424&218&158\\
	    \hline
	    WL2-3&End Device&848&436&316\\
	    \hline
	    WL2-4&End Device&1696&872&632\\
	    \hline
	    WL2-5&End Device&3392&1744&1264\\
	    \hline
	    WL2-6&End Device&6784&3488&2528\\
	    \hline
	    WL3-1&Edge Server&3115&2931&3618\\
	    \hline
	    WL3-2&Edge Server&6230&5862&7236\\
	    \hline
	    WL3-3&Edge Server&12460&11724&14472\\
	    \hline
	    WL3-4&Edge Server&24920&23448&28944\\
	    \hline
	    WL3-5&Edge Server&49840&46896&57888\\
	    \hline
	    WL3-6&Edge Server&99680&93792&115776\\
	    \hline
	\end{tabular}
\end{table}

\subsection{Measured Response Time for Single Workload}
\begin{figure}[htbp]
\flushleft
\subfigure[Short-of-breath alerts.]{
\begin{minipage}[t]{1 \linewidth}
\centering
\includegraphics[width=3.5in]{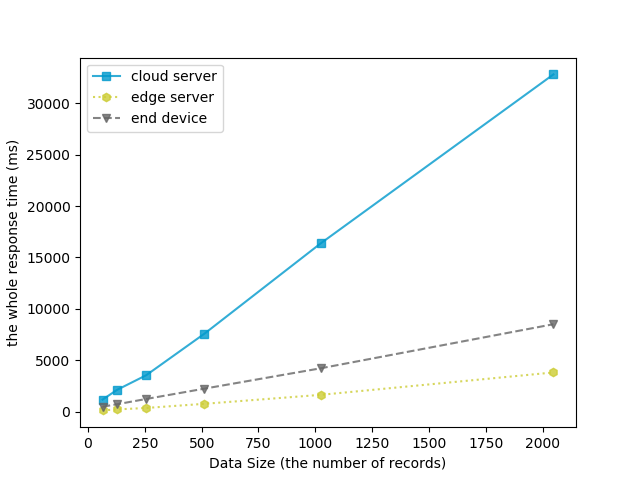}
\end{minipage}
}
\subfigure[Life-death prediction.]{
\begin{minipage}[t]{1\linewidth}
\centering
\includegraphics[width=3.5in]{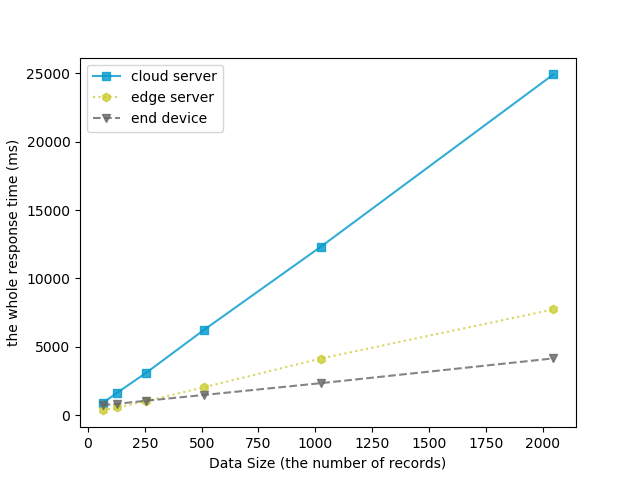}
\end{minipage}
}
\subfigure[Patient phenotype classification.]{
\begin{minipage}[t]{1\linewidth}
\centering
\includegraphics[width=3.5in]{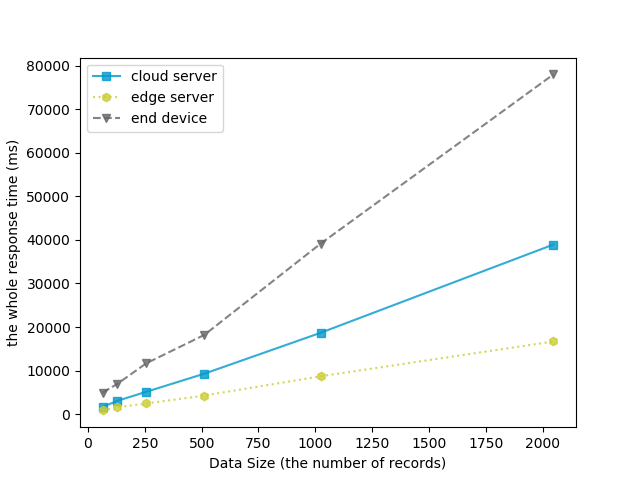}
\end{minipage}
}
\centering
\caption{The experiment response time results of each workload deployed on different layers.}
\label{deployment_exp}
\end{figure}

For validating the effectiveness of the computational results, we deploy each workload respectively on each layer to conduct the inference experiment on the real experimental environment. Then we get the real response time for each workload deployed on different layers. Figure~\ref{deployment_exp} shows the experimental results.

Figure~\ref{deployment_exp}a shows the results for the short-of-breath application. Deploying the workload on the edge server can get the lowest response time, and deploying on the cloud server can get the highest response time. Figure~\ref{deployment_exp}b shows for life-death prediction, the end device is the optimal deployment layer. And figure~\ref{deployment_exp}c shows the edge layer is the optimal deployment layer for patient phenotype classification.

By comparing table~\ref{estimated} with figure~\ref{deployment_exp}, we find most of the estimated deployment layer will get the lowest response time. For WL2-1 and WL2-2, it's better to deploy on edge server but we predict to deploy on end device. But the response time of deploying on the edge server and end devices are very close in these two workloads. 

In general, the experimental result demonstrates the effectiveness of our optimal strategy for the single healthcare workload deployment.

\begin{figure}[ht]
\centering
\includegraphics[scale=0.6]{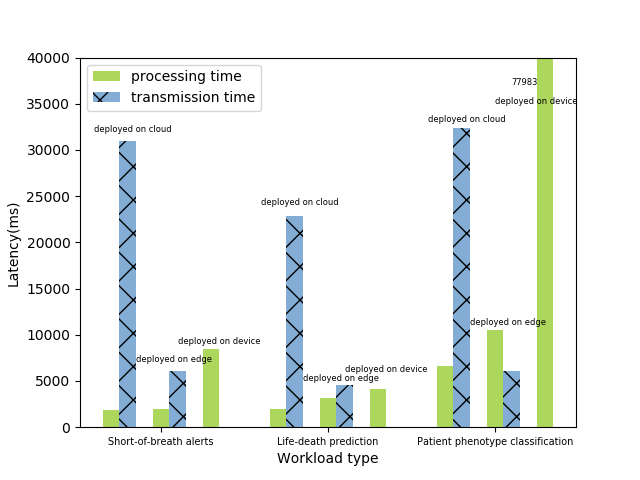}
\caption{Response time breakdown of each workload.}
\label{breakdown}
\end{figure}

What's more, we begin exploring the critical influence factor of the response time by the breakdown figure~\ref{breakdown}. We choose the workload WL1-6, WL2-6, and WL3-6 to represent the three applications.  Figure~\ref{breakdown} shows the processing time and transmission time of the three workloads.

From figure~\ref{breakdown}, we have the following observations. 

The model of patient phenotype classification is more complicated than the other two applications. So the processing time on the end device is respectively higher. Therefore, the transmission time has a smaller influence on this condition.

For life-death prediction, the optimal deployment layer is the end device. Because the number of parameters of this model is small, the end device can satisfy the requirements of this model. In this situation, workloads don't need to be offloaded on the edge or cloud server.

We can conclude that the more simplified the workload models, the greater influence the transmission time has. Therefore, computing near the user may get the lowest response time.

On the contrary, for a heavy-weight workload, being processed on the higher layer may get the lowest response time.

So for the AI workload deployment on edge computing framework, we need to estimate the computation ability of the devices on different layers and the network condition firstly. Then we can decide which layer to offload the workload by trading off between the processing time and transmission time.

\subsection{Multi-job Scheduling Strategy}
To validate the scheduling algorithm, we extract 10 jobs from the above experimental workload execution time results. And we normalize their response time. Meanwhile, we set release time for each workload to simulate the real-world environment. Tabel~\ref{schedule} shows the scheduling experiment setting.

\begin{table*}[htbp]
	\centering
	\caption{Processing time, transmission time and release time of jobs deployed on different layer. } 
	\label{schedule} 
	\begin{tabular}{|c|c|c|c|c|c|c|c|}  
	   \hline
	   \multicolumn{1}{|c|} {\multirow{2}{*}{ \shortstack{Job No.}}}
	       &   
	    \multicolumn{1}{c|}{\multirow{2}{*}{ \shortstack{Release}}}
	    &   
	    \multicolumn{1}{c|}{\multirow{2}{*}{Priotity Weight}}
	   &
	   \multicolumn{2}{c|} {Deployed on Cloud Server} 
	   	     &
	   \multicolumn{2}{c|} {Deployed on Edge  Server} 
	   &
	   Deployed on End Device\\
	    \cline{4-8}
	    &&&Processing&Transmission
	    &
	    Processing &Transmission
	    &
	    Processing \\
	    \hline
	   J1&1&2&6&56&9&11&14\\
	    \hline
	   J2&1&2&3&32&3&6&12\\
	      \hline
	   J3&3&1&4&12&6&2&49\\
	   	   \hline
	   J4&5&1&7&23&11&5&69\\
	   	   \hline
	   J5&10&2&4&27&5&5&11\\
	   	   \hline
	   J6&20&2&5&70&5&14&22\\
	   \hline
	   J7&21&2&5&70&5&14&22\\
   \hline
	   J8&21&1&4&12&6&2&49\\
	   \hline
	   J9&22&1&4&12&6&2&49\\
	   \hline
	   J10&25&1&7&23&11&5&69\\
	    \hline

	    \hline
	\end{tabular}
\end{table*}

We use algorithm 2 to calculate the efficient workload deployment strategy for these 10 jobs. Firstly we calculate the initial feasible deployment strategy. We choose the best deployment layer for each workload by time sequence. Then we adjust the deployment strategy using the heuristic method iteratively. 

\begin{figure}[ht]
\centering
\includegraphics[scale=0.4]{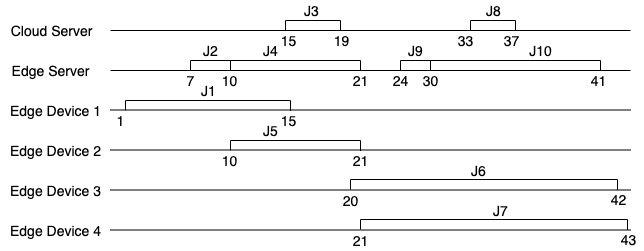}
\caption{Allocation Strategy Using Algorithm 2.}
\label{sche_fig}
\end{figure}

Figure~\ref{sche_fig} shows the scheduling method by our workload allocation algorithm 2. The horizontal axis represents the time sequence. And it shows the start execution time and completion time of each job on the deployment layer. Our workload allocation strategy gets 150 whole response time and the last completion time is 43. And there are 4 workloads that need to be deployed on end devices, 4 on the edge server, and 2 on the cloud server. And we can see the deployment layer of each job is not optimal for the single workload.

For comparison, we calculate the response time of the other 4 deployment strategies. We deploy all the workloads on the cloud, edge server, end devices, or each job's optimal layers. 

\begin{figure}[ht]
\centering
\includegraphics[scale=0.38]{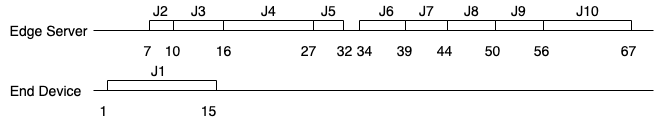}
\caption{Allocation Strategy Using the Optimal Layer for Each Job.}
\label{opt_fig}
\end{figure}

And figure~\ref{opt_fig} shows the scheduling strategy by choosing the optimal deployment layer of each job. For job 1, the optimal deployment layer is the cloud server. And for other jobs are edge server. We can see from figure~\ref{opt_fig} that a lot of jobs need to wait for the completion of the last one. So it leads to great delays, which indicates the necessity of our strategy.

\subsection{Measure Total Response Time of Multi Jobs}
We calculate the sum of whole response time and the last completion time of our allocation and scheduling strategy and other 4 strategies. Table~\ref{comparison} shows the results of these strategies.

\begin{table}[htbp]
	\centering
	\caption{Reponse Time Using Different Algorithms. } 
	\label{comparison} 
	\begin{tabular}{|p{3.4cm}|p{2cm}|p{2cm}|}  
	   \hline
	   Strategy&Whole Response Time&Last Response Time\\
	    \hline
	    Our Allocation Strategy&150&43\\
	    \hline
	    Deployed on the Optimal Layer for Each Job&227&67\\
	    \hline
	    Deployed on Cloud Server&291&74\\
	    \hline
	    Deployed on Edge Server&416&100\\
	    \hline
	    Deployed on End Device&366&94\\
	    \hline
	\end{tabular}
\end{table}

Our strategy has the lowest whole response time and the lowest last response time. And for the whole response time, our optimal deployment strategy respectively gets 33\%, 48\%, 63\%, and 59\% lower response time than other strategies. The comparison results demonstrate our strategy can get the minimum whole response time and last response time.

\section{Related Work}
In this section, we review the recent and related work of edge computing workload deployment and resource allocation.

Xu et al.~\cite{xu2017zenith} proposed a resource allocation model in the edge computing platform. Cao et al.~\cite{cao2019optimal} proposed a task allocation strategy to reduce resource consumption. However, these studies focus on resource allocation among edge computing servers, assuming that the workload is deployed on the edge computing layer.

Fan et al.~\cite{fan2018application} designed an application-aware workload allocation scheme to minimize the response delay. This scheme decides which cloudlet to allocate the workload by considering the computing resources allocated location and the type of the application.

Chen et al.~\cite{chen2018data} proposed an optimal caching strategy for mobile services. But this work doesn't consider the multiple jobs deployment problem.

Feng et al.~\cite{feng2019computation} proposed an optimal offload algorithm to maximize the data utility and minimize energy consumption. It encourages offload computing tasks to the MEC server and doesn't consider the other layer to offload the computing tasks.

Dong et al.~\cite{dong2019computation} proposed a graph cut problem solution to reduce the transmission consumption and energy consumption during offloading.

Chen et al.~\cite{chen2019data} focused on task allocation by considering the importance of the task. They proposed an allocation approach to accelerate the computational efficiency.

To the best of our knowledge, this paper is the first research effort focusing on the AI workload allocation and multiple jobs scheduling strategy in time sequence towards the three-layer cloud/edge/device hierarchical framework.

\section{Conclusions}

In this work, we focus on the AI-oriented medical workload allocation algorithm on cloud/edge/device computing hierarchically-structured framework. Our work can be applied to reduce the response time for latency-sensitive workload such as ICU patient monitor applications.

Firstly, we propose a method to deploy on the optimal layer for a single workload. We analyze the complexity of the model, the device computation ability of the three edge computing layer and the network condition. Then we estimate the response time by summing up the processing time and transmission time. Therefore, we can choose the optimal deployment layer.

Based on the above algorithm, we also propose an allocation and scheduling strategy for multiple jobs in the time sequence. This strategy's goal is to minimize the whole response time of all jobs, which considers the different priorities of these jobs. We use a heuristic algorithm in this strategy.

At last, we have conducted several experiments for medical AI applications on the cloud/edge/device environment. And we choose three latency-sensitive healthcare applications: short-of-breath alerts, life-death prediction, and patient phenotype classification. And we use the real-world medical datasets. 

Experimental results show that the allocation strategy will greatly influence the response time of the workload. And by comparison with other strategies, the results demonstrate the effectiveness and efficiency of our workload allocation strategies.

\section{Acknowledgements}
This research is partially supported by the Shenzhen Institute of Artificial Intelligence and Robotics for Society (AIRS) and the State Key Laboratory of Computer Architecture, Institute of Computing Technology, Chinese Academy of Sciences.
\bibliographystyle{plain}
\bibliography{IEEEexample}

\end{document}